\documentclass[12pt]{article}
\usepackage[dvips]{graphicx}
\usepackage[dvips]{graphics}
\newcommand{\newc}{\newcommand}
\newc{\lra}{\leftrightarrow}
\newc{\beq}{\begin{equation}}
\newc{\eeq}{\end{equation}}
\newc{\barr}{\begin{eqnarray}}
\newc{\earr}{\end{eqnarray}}
\title{The effects of deformation and pairing correlations on
nuclear charge form factor}

\author{
Ch.~C.~Moustakidis$^1$,
T.~S.~Kosmas$^{1,2}$,
F.~Simkovic$^{1,3}$ and
Amand~Faessler$^1$
\\
{\it  $^{1}$Institute of Theoretical Physics, University of Tuebingen,\qquad \qquad \qquad } \\
{\it D-72076 Tuebingen, Germany \qquad \qquad } \\
{\it $^{2}$Theoretical Physics Division, University of Ioannina, \qquad \qquad \qquad  } \\
{\it GR-45110 Ioannina, Greece \qquad \qquad \qquad \qquad}\\
{\it $^{3}$Department of Nuclear Physics, Comenius University, \qquad \qquad \qquad  } \\
{\it SK-84215 Bratislava, Slovakia \qquad \qquad }  }

\date{}
\begin{document}

\maketitle

\begin{abstract}
A set of moderately deformed $s-d$ shell nuclei is employed for
testing the reliability of the nuclear ground state wave functions
which are obtained in the context of a BCS approach and offer a
simultaneous consideration of deformation and pairing correlations
effects. In this method, the mean field is assumed to be an
axially symmetric Woods-Saxon potential and the effective two-body
interaction is a monopole pairing force. As quantities of main
interest we have chosen the nuclear form factors, the occupancies
of the active (surface) orbits and the Fermi sea depletion, which
provide quite good tests for microscopic descriptions of nuclei
within many body theories. For our comparisons with results
emerging from other similar methods, an axially deformed harmonic
oscillator field is also utilized.
\end{abstract}

%\noindent Keywords: LSP, Direct neutralino search, ionization
%electrons, electron detection,
% dark matter, WIMP.\\
\

\section{Introduction.}

In recent nuclear structure calculations, the reliability of
various predictions associated with phenomena of several prominent
electroweak processes ($\beta$-decay modes, lepton-capture by
nuclei, neutrino-nucleus reactions, etc.) relies upon the accurate
description of the nuclear many-body wave functions
\cite{Suho-Civi,Sim-Fes}. Among the nuclear structure aspects
which can quite strongly influence such predictions, the
consideration of the deformation
\cite{Nojarov,Sarrigu01,Sim-Mous,Kosmas03} and the nucleon-nucleon
correlations effects \cite{Antonov,Kosmas01,Papacon,Mas-Mous}
(wherever they might be appreciable) is of significant importance.
As it is well known, the ground state nucleon-nucleon correlations
\cite{Antonov} are of particular significance to be reliably
incorporated especially when investigating processes for which the
branching ratios or decay (event) rates are highly suppressed and,
therefore, the corresponding transition matrix elements are very
sensitive \cite{Sim-Mous,Kosmas01}. Moreover, the nuclear
deformation, e.g. for axially symmetric nuclear systems, should
necessarily be considered by applying appropriate treatments and
using reliable methods \cite{Nojarov,Sarrigu01,Sim-Mous,Kosmas03}.

The main purpose of the present work is to address the effects of
deformation and pairing correlations on  a few nuclear observables
of the ground state for moderately deformed s-d shell nuclei in
the spirit of approaches based on BCS quasi-particles. The
properties of main interest are the ground state charge form
factor $F_{ch}({\bf q})$
\cite{Sinha73,Vries,Sound88,Wessel97,Bender03}, as well as the
occupancy (vacancy) numbers of several orbitals around the Fermi
level \cite{McNeil90,Kosmas92,Lapikas,Pandharipande97,Brown03}.
Such quantities (also those referred to low-lying excitations like
transition densities, transition form factors, etc.) constitute
very common tests in nuclear structure theory for the many-body
model wave functions \cite{Suho-Civi,McNeil90}. Experimentally,
the above properties are known with very high accuracy ($\approx
1\%$), e.g. from experiments of elastic scattering of electrons,
protons, etc. off target nuclei
\cite{Sinha73,Vries,Sound88,Wessel97}. On the theoretical side,
average potentials (spherical Woods-Saxon or harmonic oscillator,
etc.) employed in BCS theories and nucleon orbitals, both of which
are directly related to gross nuclear features of the ground
states, have been quite effective in describing the dynamics of
many mainly spherical nuclei
\cite{Mas-Mous,Kosmas92,Brown03,Schwi}. However, even though the
direct calculation of the charge form factor (and other gross
nuclear properties) through a nuclear model ground state (g.s.)
wave function is a very extensively discussed issue, the
reproducibility of the scattering data in the high momentum
transfer region ($q>2-3$ fm$^{-1}$) with the experimental accuracy
has not yet been achieved (in the context of any nuclear model)
for most of the nuclear systems.

The existence of nuclear deformation, in general, is closely
related to the non-spherical components of the proton (neutron)
density distribution and the corresponding proton (neutron)
nuclear form factor whereas the nucleon-nucleon correlations are,
to a large extent, related to the nuclear surface texture, the
main characteristics of which are the occupation probabilities of
the active orbits. These quantities are studied in single-nucleon
transfer reactions \cite{Lapikas}, but often the results differ
substantially from the predictions of independent particle
approximations and other methods, due to the fact that, in many of
these models, the correlations and deformation phenomena are not
(or may not correctly \cite{Van-Neck}) taken into account
\cite{Antonov}. Hence, details of the nuclear surface and also
bulk phenomena for many nuclei may have not correctly been
understood in terms of many existing models.

The present method employs a BCS ground state built out of a
deformed WS basis wave functions. The required two-body
interaction is taken to have a schematic-type, the monopole part
of which is scaled phenomenologically (separately for protons and
neutrons) so as to be consistent with the observed proton and
neutron separation energies within the BCS procedure
\cite{Sim-Mous,Madland88}. At first, we study the nuclear form
factors for the set of deformed nuclei $^{24}_{12}$Mg,
$^{26}_{12}$Mg, $^{28}_{14}$Si and $^{32}_{16}$S (s-d shell
nuclei), by considering them as systems having axially symmetric
$\beta$ (quadrupole) deformation only \cite{Moller,Lalazi}. They
are rather well deformed isotopes with the following values of the
parameter $\beta$: $\beta (^{24}$Mg$)=0.416$ \cite{Lalazi} (0.374
\cite{Moller}), $\beta (^{26}$Mg$)=0.296$ \cite{Lalazi} (-0.310
\cite{Moller}), $\beta (^{28}$Si$)=-0.328$ \cite{Lalazi} (-0.478
\cite{Moller}) and $\beta (^{32}$S$)=0.186$ \cite{Lalazi} (0.000
\cite{Moller}). In the present work we apply also values of the
deformation parameter $\beta$ which originated from experimental
values of the quadrupole moment, taken from Ref. \cite{Stone01}

We next proceed with the study of the effects of deformation and
pairing interactions on the surface features of these nuclei,
namely, on the smearing of their Fermi surface which is very
sensitive to the orbit occupancies of the active space. In our
method, deformation effects and pairing correlations for
proton-proton and neutron-neutron interactions may simultaneously
and in a quite reliable manner be considered which is in accord
with the experimental indications of Ref. \cite{Wessel97} showing
that these effects must be treated as much as possible on equal
footing. Finally, for the sake of comparison with the results of
other similar methods, we study the same properties for the above
isotopes, but now by expanding the required deformed Woods-Saxon
wave functions into a deformed harmonic oscillator (DHO) basis.

We remark that, spherical WS potentials parameterized for the well
depth, skin thickness, radius and spin-orbit constants, etc. as in
Ref. \cite{Tanaka79}, have been used by us in the past for
microscopic random phase approximation (RPA) calculations of decay
(event) rates in electroweak processes ($\beta\beta$-decay
\cite{Sim-Fes}, flavor violating processes \cite{Kosmas01,Schwi},
etc.) for various nuclei. Several of these calculations must be
extended to non-spherical systems or must be corrected to include
deformation and/or correlations effects. The present method offers
some advantages for such purposes \cite{Sarrigu03,Sim-Pace}.

In the remainder of the paper, we first describe the main
ingredients of our method (Sect. 2), then we present the results
obtained and compare them with existing experimental data (Sect.
3), and finally (Sect. 4) we summarize the main conclusions
extracted.

\section{Brief description of the formalism  }

\subsection{Axially symmetric Harmonic Oscillator and Wodds-Saxon bases}
For the description of axially symmetric nuclear systems, where
the cylindrical coordinates $(r,z,\phi)$ are appropriate, a
deformed harmonic oscillator (DHO) potential has been primarily
used \cite{Nillson}. The eigenfunctions of this unharmonic
oscillator are described in terms of two frequencies:
$\omega_{z}$, in the direction of the z-axis, and
$\omega_{\perp}$, for the motion perpendicular to the symmetry
z-axis. Then, the space-spin DHO single particle wave functions
are written as \cite{Nillson}
\begin{equation}
\Psi_{n_\rho,n_{z},\Lambda,\Sigma} ({\bf r},\mbox{\boldmath
$\sigma$}) \, \equiv \, \Psi_{n_\rho,n_{z},\Lambda,\Sigma}
(\rho,z,\phi,\mbox{\boldmath $\sigma$}) \, = \,
\psi_{n_\rho}^{\mid \Lambda \mid}(\rho) \ \psi_{n_{z}} (z) \
\psi_{\Lambda}(\phi) \ \chi(\Sigma) \label{DHO-WF}
\end{equation}
where $\Sigma$ and $\Lambda$ are the projections on the z-axis of
the spin and orbital angular momentum, respectively.
$\chi(\Sigma)$ stands for the spin wave function and the
components of the spatial parts are defined as follows:

(i) The radial part is usually written in terms of a dimensionless
coordinate $\eta$ defined by
\begin{equation}
\eta=R_{\perp}^2 \rho^2, \qquad R_{\perp}=(\frac{m
\omega_{\perp}}{\hbar})^{\frac{1}{2}},
\end{equation}
($m$ is the nucleon mass) as
\begin{equation}
\psi_{n_\rho}^{\mid \Lambda \mid}(\rho) =
N_{n_\rho}^{\mid\Lambda\mid}
(\frac{2m\omega_{\perp}}{\hbar})^{\frac{1}{2}}
\eta^{\frac{\mid\Lambda\mid}{2}} e^{-\frac{\eta}{2}}
L_{n_\rho}^{\mid \Lambda\mid}(\eta), \label{radial}
\end{equation}
where
\begin{equation}
N_{n_\rho}^{\mid\Lambda\mid} = (\frac {n_{\rho}
!}{(n_{\rho}+\mid\Lambda\mid)!})^{\frac{1}{2}},
\end{equation}
is a normalization factor and $L_{n_\rho}^{\mid \Lambda
\mid}(\eta)$ are the associated Laguerre polynomials.

(ii) The z-dependent part is similarly written in terms of a
dimensionless variable $\xi$
\begin{equation}
\xi=R_{z} z, \qquad R_{z}= (\frac{m
\omega_{z}}{\hbar})^{\frac{1}{2}}, \label{Rz-par}
\end{equation}
as
\begin{equation}
\psi_{n_{z}} (z) = N_{n_z}
(\frac{m\omega_{z}}{\hbar})^{\frac{1}{4}} e^{-\frac{\xi^2}{2}}
H_{n_z}(\xi)
\end{equation}
where again
\begin{equation}
N_{n_z} = (\sqrt{\pi} 2^{n_z} n_z !)^{-\frac{1}{2}},
\end{equation}
is a normalization factor and $H_{n_z}(\xi)$ are the Hermite
polynomials.

(iii) the $\phi$-dependent part is given by
\begin{equation}
\psi_{\Lambda}(\phi)=\frac{1}{\sqrt{2 \pi}} e^{i \Lambda \phi},
\label{phi-dep}
\end{equation}
which is normalized to unity.

The coordinate scaling parameters $R_{\perp}$ and $R_z$, which
have dimensions of inverse length, characterize the motion in the
perpendicular and z-direction, respectively. Obviously, the level
of anisotropy of the field is represented by the difference
between the two frequencies, $\omega_{\perp}$ and $\omega_{z}$,
which are written in terms of the usual quadrupole deformation
parameter $\beta$ as \cite{Nillson}
$$
\omega_{z}=\omega_{0}(\beta)\left(1-\frac{2}{3}\beta \right),
\qquad  \qquad
\omega_{\perp}=\omega_{0}(\beta)\left(1+\frac{1}{3}\beta \right)
$$
where $\omega_{0}$ is weakly dependent on $\beta$ (enough to
conserve the nuclear volume). From the latter two equations we
obtain
$$
\beta=(\omega_{\perp}-\omega_{z})/ \omega_{0}.
$$
The parameter $\beta$ is defined so that $\beta > 0$ correspond to
the so-called prolate shapes and $\beta <0 $ correspond to oblate
shapes.

The eigenfunctions of the Hamiltonian involving an axially
symmetric Woods-Saxon potential have a good quantum number $K$,
the projection of the total angular momentum, ${\bf J} = {\bf L} +
{\bf S}$, on the symmetry z-axis, i.e. $K=\Lambda+\Sigma$ (due to
the assumed symmetry they are eigenfunctions of the ${\hat J}_z$
operator). These eigenfunctions can be expanded into basis
functions like those of Eq. (\ref{DHO-WF}), namely, into the
eigenfunctions of a deformed harmonic oscillator potential. In the
intermediate region of the parameter $\beta$ we are interested in,
$0.15 \le \beta \le 0.50$, one could employ either Eq.
(\ref{DHO-WF}) or total angular momentum coupled bases like the
$\Psi (N,n_z,\Lambda,K)$ (see Ref. \cite{Nillson} p. 118) which we
apply in this work (here the principal quantum number $N$ is
defined as $N=n_z+2n_{\rho}+\mid \Lambda \mid$). In this case the
expansion of the WS basis takes the form
\begin{eqnarray}
\Phi_i
(K)&=&\sum_{N,n_z,\Lambda}\alpha_i(N,n_z,\Lambda)\Psi(N,n_{z},\Lambda,\Sigma=1/2,K)
\nonumber\\
&+& \sum_{N',n_z',\Lambda'}\alpha_i'(N',
n'_z,\Lambda')\Psi(N',n'_z,\Lambda'=\Lambda+1, \Sigma'=-1/2,K)
\label{wf-WS}
\end{eqnarray}
where $\alpha_i$ and $\alpha_i'$ are expansion coefficients
obtained from the diagonalization procedure. The time-reversed
partners of the states (\ref{wf-WS}) have the form
\begin{eqnarray}
\tilde{\Phi}_i
(K)&=&\sum_{N,n_z,\Lambda}\alpha_i(N,n_z,\Lambda)\Psi(N,n_{z},-\Lambda,\Sigma=-1/2,-K)
\nonumber\\
&-&
\sum_{N',n'_z,\Lambda'}\alpha_i'(N',n'_z,\Lambda')\Psi(N',n'_z,\Lambda'=-\Lambda-1,
\Sigma'=1/2,-K) \label{wf-WS-trev}
\end{eqnarray}
Using Eqs. (\ref{wf-WS}) and (\ref{wf-WS-trev}), we can write the
expression for the point proton density distribution in an axially
symmetric nucleus as
\begin{equation}
\rho ({\bf r})=\sum_{i} V_{pi}^2 [\Phi^{*}_i(K) \Phi_i (K)+
\tilde{\Phi}^{*}_i(K) \tilde{\Phi}_i(K)] \label{density}
\end{equation}
where $V_{pi}^2$ stands for the occupation probability of the
$i^{th}$ proton state (in the deformed basis), determined from the
solutions of the BCS equations. The proton density distribution,
$\rho ({\bf r})$, Eq. (\ref{density}), is related to the proton
form factor $F({\bf q})$ of the nucleus in question by the well
known expression (Born approximation)
\begin{equation}
F ({\bf q})=\int \rho ({\bf r}) e^{i{\bf q}{\bf r}} d{\bf r}
\label{FF-1}
\end{equation}
By choosing ${\bf q}$ in the direction of ${\hat z}$ we have,
$i{\bf q}{\bf r}=iqr\cos{\theta}=iqz$. Under this assumption, the
nuclear form factor, Eq. (\ref{FF-1}), can be written in the form
\begin{equation}
F (q)=2 \sum_i \left( \sum_{N,n_z,\Lambda} \sum_{N',n_z'}
\alpha_i(N,n_z,\Lambda) \ \alpha_i(N',n_z',\Lambda ) \
I(n_z,n_z',q) \ \delta_{n_{\rho},n_{\rho}'} \ \right) \  V_{pi}^2
\label{FF-2}
\end{equation}
where
\begin{equation}
I(n_z,n_z',q)= \int_{-\infty}^{\infty} \psi_{n_{z}}^{*} (z)
\psi_{n_{z}'} (z)e^{iqz} dz . \label{FF-3}
\end{equation}
The first part of our calculations in the present work (see Sect.
3) relies upon Eqs. (\ref{FF-2}) and (\ref{FF-3}).

\subsection{The BCS model}
The ground state of even-even deformed nuclei is determined by the
deformed pairing Hamiltonian, which includes monopole
proton-proton and  neutron-neutron pairing interactions as
\cite{Sim-Mous,Soloviev}
\begin{eqnarray}
H&=& \sum_{s}(\epsilon_{p s}^{0}-\lambda_p) \sum_{\sigma} c_{p
s\sigma}^{\dagger}   c_{ps \sigma}+ \sum_{s}(\epsilon_{n
s}^{0}-\lambda_n) \sum_{\sigma} c_{ns\sigma}^{\dagger}   c_{ns
\sigma}
\nonumber\\
&&-G_{pp} \sum_{s,s'} {S^{\dagger}_{spp}} S_{s'pp} - G_{nn}
\sum_{s,s'} {S^{\dagger}_{snn}} S_{s'nn}  \ , \label{eq:1}
\end{eqnarray}
where $\epsilon_{p s}^{0}$ and $\epsilon_{n s}^{0}$ are the
un-renormalized single particle energies for protons and neutrons,
respectively. $\lambda_p$ ($\lambda_n$) is the proton (neutron)
Fermi energy and the operator ${S^{\dagger}_{s\tau \tau'}}$
creates pairs in time reversed orbits written as
\begin{equation}
{S^{\dagger}_{spp}}= \sum_\sigma c_{ps\sigma}^{\dagger} c_{ps
\tilde{\sigma}}^{\dagger}\ , \qquad {S^{\dagger}_{snn}} =
\sum_\sigma c_{ns\sigma}^{\dagger} c_{ns \tilde{\sigma}}^{\dagger}
\ . \label{eq:2}
\end{equation}
Here, $c_{\tau s\sigma}^{\dagger}$ and $c_{\tau s\sigma}$ stand
for the creation and annihilation operators of a particle
($\tau=p$ and $\tau=n$ denote proton and neutron, respectively) in
the axially symmetric Woods-Saxon potential. $\sigma$ is the sign
of the angular momentum projection $K$, namely, $\sigma=\pm 1$. We
note that the intrinsic states are twofold degenerate, since the
states with $K$ and $-K$ have the same energy as a consequence of
the time reversal invariance (the symbol $\sim$ indicates time
reversed states). The pairing interaction strengths $G_{pp}$ and
$G_{nn}$, which characterize the associated monopole interaction
for proton or neutron pairs, respectively, are determined by
fitting the theoretical pairing gaps to the average empirical ones
parameterized as, $\Delta_{pp}= \Delta_{nn}=12.84/A^{1/2}$ MeV
\cite{Madland88}, where $A$ represent the mass number of the
isotope in question.

\section{Results and Discussion  }

The first part of the present calculations refers to the proton
form factors of the deformed nuclei $^{24}_{12}$Mg,
$^{26}_{12}$Mg, $^{28}_{14}$Si and $^{32}_{16}$S carried out as
follows. At first, in the spirit of the independent particle
approximation (Slater determinant calculations), we used a
deformed Woods-Saxon field parameterized as in Ref.
\cite{Tanaka79} (in this step there is no correlations effect). In
the second step, the pairing correlations are inserted in our
calculation by means of the BCS theory, namely through the
monopole pairing interaction, a schematic type nucleon-nucleon
force. The strength of this pairing force, is determined
separately for protons and neutrons as in Ref. \cite{Sim-Mous}
(see Sect. 2.2). In the third step of the calculational procedure,
for comparison, we evaluated the same proton form factors
utilizing now a deformed harmonic oscillator basis in the BCS
ground state, as discussed in Sect. 2. Thus, the latter results
include correlations of the same kind as those of the second step.
For all the cases studied, we adjusted the parameter $R_z$ of Eq.
(\ref{Rz-par}) so as, for each isotope, the first diffraction
minimum of the nuclear form factor to appear at the experimental
position (value of the momentum transfer $q$). For the deformation
parameter $\beta$, throughout this part of work we used the more
recent calculations of Ref. \cite{Lalazi}, which are, generally,
in agreement with the predictions of the macroscopic-microscopic
model of Ref. \cite{Moller}.

%Fig1
\begin{figure}[h]
 \centering
 \includegraphics[height=13.5cm,width=13.5cm]{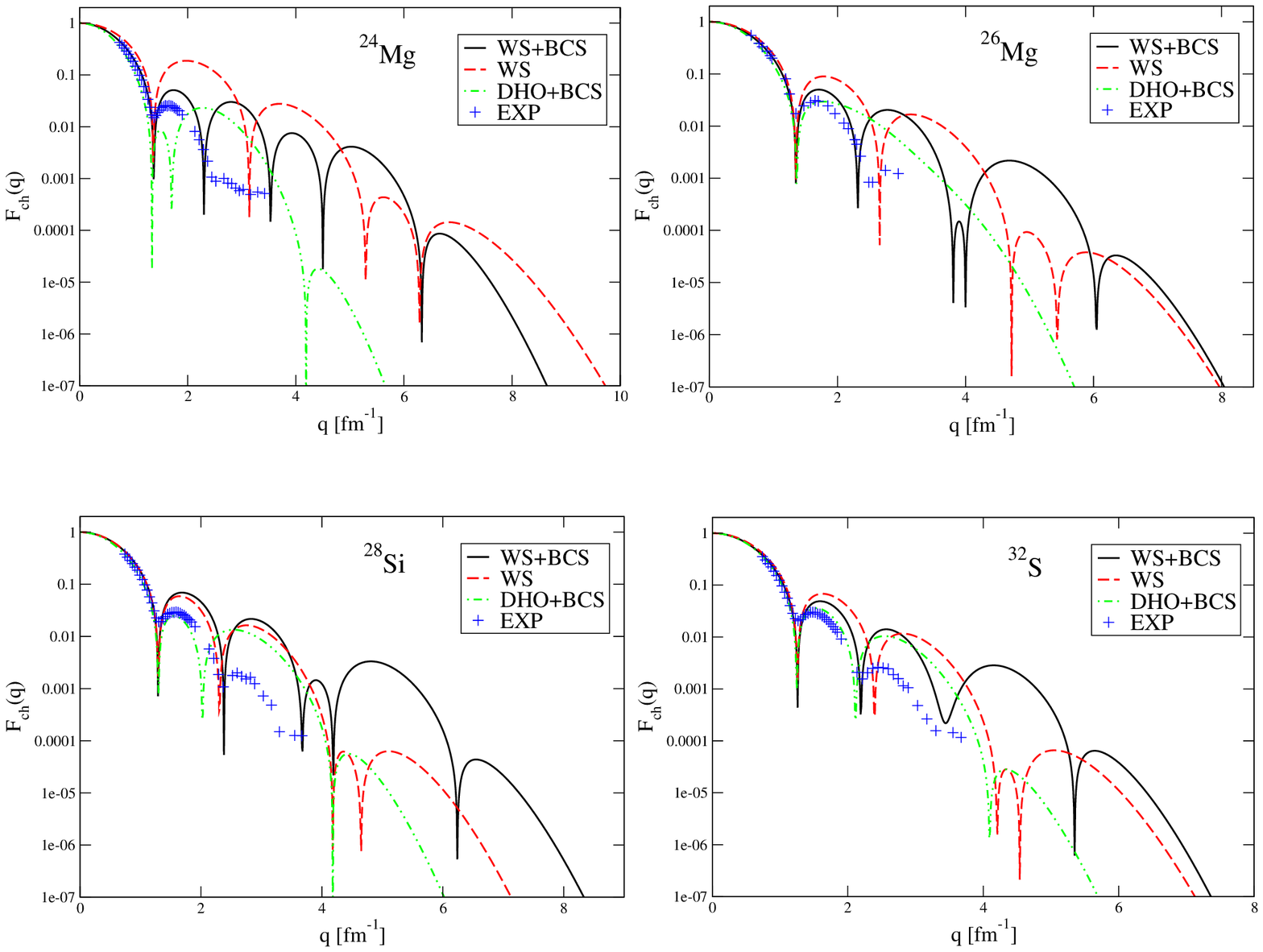}
 %\hspace{0.5cm}
  \caption{ Calculated and experimental
\protect\cite{Sinha73,Vries,Sound88,Wessel97} charge form factors
for the deformed nuclei $^{24}_{12}$Mg, $^{26}_{12}$Mg,
$^{28}_{14}$Si and  $^{32}_{16}$S. The results labeled WS+BCS
(solid line) include pairing correlations and deformation effects
while those labeled WS (dashed line) originate from a simple
Slater determinant constructed out of a deformed Woods-Saxon basis
wave functions. The dashed dotted curve, DHO+BCS, comes from a
deformed harmonic oscillator field employed in the BCS ground
state.}
\end{figure}

The results obtained as described above, are presented in Fig. 1,
where we use the notation WS to denote the deformed (axially
symmetric) Woods-Saxon results, the notation WS+BCS to represent
the results containing in addition pairing correlations, and the
notation DHO+BCS for the results which consider the deformation
and pairing-correlations effects through the harmonic oscillator
model. The main conclusions stemming out of the above calculations
are the following. As can be seen from Fig. 1, the incorporation
of pairing correlations (second step) via the BCS procedure,
affects appreciably the shape of the charge form factors (compare
the plots corresponding to WS and WS+BCS results), especially for
$q>2$ fm$^{-1}$, which is a well known outcome. Furthermore, the
simultaneous consideration of deformation and pairing correlations
in the BCS ground state with Woods-Saxon potential, reproduces the
experimental form factor data (and charge density distribution)
much better than the deformed harmonic oscillator model (compare
results labeled WS+BCS and DHO+BCS). As it is evident from Fig. 1,
nearly all the diffraction minima of the results labeled WS+BCS
are produced quite close to their experimental momentum transfer
locations (also some additional minima not observed by experiments
up to now are predicted for values of $q > 3-4$ fm$^{-1}$)
\cite{Brown03}. For high momentum transfer (close to the nuclear
center of the charge density distribution) the reproducibility
becomes worse due to the presence of other sorts of correlations
not included in our results (short range nucleon-nucleon
correlations, etc.) and other kinds of effects (meson-exchange
corrections, etc.).

%%%%%%%%%%%%%%%%%%%%%%%%%%%%%%%%%%%%
We investigate also the effect of the deformation on charge form
factors as well the occupancy numbers by using various values of
the deformation parameter $\beta$ which originated from Ref.
\cite{Stone01} (see also Table 1). More specifically the parameter
$\beta $ calculated through the relation
\begin{equation}
\beta=\sqrt{\frac{\pi}{5}}\frac{Q_0}{Z r_c^2} \label{beta-1}
\end{equation}
where $Q_0$ is the intrinsic quadrupole moment (in barn), $r_c$ is
the charge radius (in fm) and $Z$ the number of the proton. $Q_0$
calculated from the experimental quadrupole moment $Q$ via the
relation
\[Q_0=-\frac{7}{2} Q \]
The relative values are exhibited in Table 1.

Fig. 2 focuses on the nucleus $^{24}$Mg and $^{32}$S showing the
calculated nuclear form factors for various values of the
deformation parameter $\beta$ where taken for Table 1. We see
that, the shape of the form factor is sensitive to changes of the
quadrupole deformation. It is worth noting that, in general, the
values of the parameter $\beta$ affect twofold the form factor
results. First, they define the set of the single particle WS wave
functions for Eq. (\ref{FF-2}) and energies $\epsilon_{p s}^{0}$
of Eq. (\ref{eq:1}) and, second, through the single-particle
energies $\epsilon_{p s}^{0}$ and the subsequently given pairing
strengths, they determine the values of the orbit occupancies
$V_{pi}^2$ entering Eq. (\ref{FF-2}).

%%%%%%%%%%%%%%%%%%%%%%%%%%%%%%%%%%%%
\vspace{0.5 cm}
\begin{table}[h]
\begin{center}
\caption {The experimental values of the quadrupole moment $Q$ and
the corresponding values of the deformation parameter $\beta$
taken from Ref. \protect\cite{Stone01}.}
\vspace{0.5 cm}
\begin{tabular}{|ccccc|}
%\toprule
\hline \hline
&   &   &   &     \\
 Nucleus & Z & $r_c$ (fm) & $Q$ (barn)  & $\beta$ \\
\hline
&   &   &   &     \\
 $^{24}$Mg & 12 &  3.126  & $-0.16 \rightarrow$ -0.29 & 0.38 $\rightarrow $ 0.69  \\
&   &   &   &    \\
 $^{26}$Mg & 12 &  3.083  & -0.11 $\rightarrow $ -0.21 & 0.27 $\rightarrow $ 0.51  \\
&   &   &   &     \\
 $^{28}$Si & 14 &  3.177  & +0.16 $\rightarrow $ +0.18 & -0.31 $\rightarrow $ -0.35 \\
&   &   &   &    \\
 $^{32}$S  & 16 &  3.282  & -0.12 $\rightarrow $ -0.18 & 0.19 $\rightarrow $ 0.29 \\
\hline \hline
% \botrule
\end{tabular}
\end{center}
\end{table}

%%%%%%%%%%%%%%%%%%%%%%%%%%%%
%Fig2
\begin{figure}[h]
 \centering
 \includegraphics[height=8.5cm,width=8.5cm]{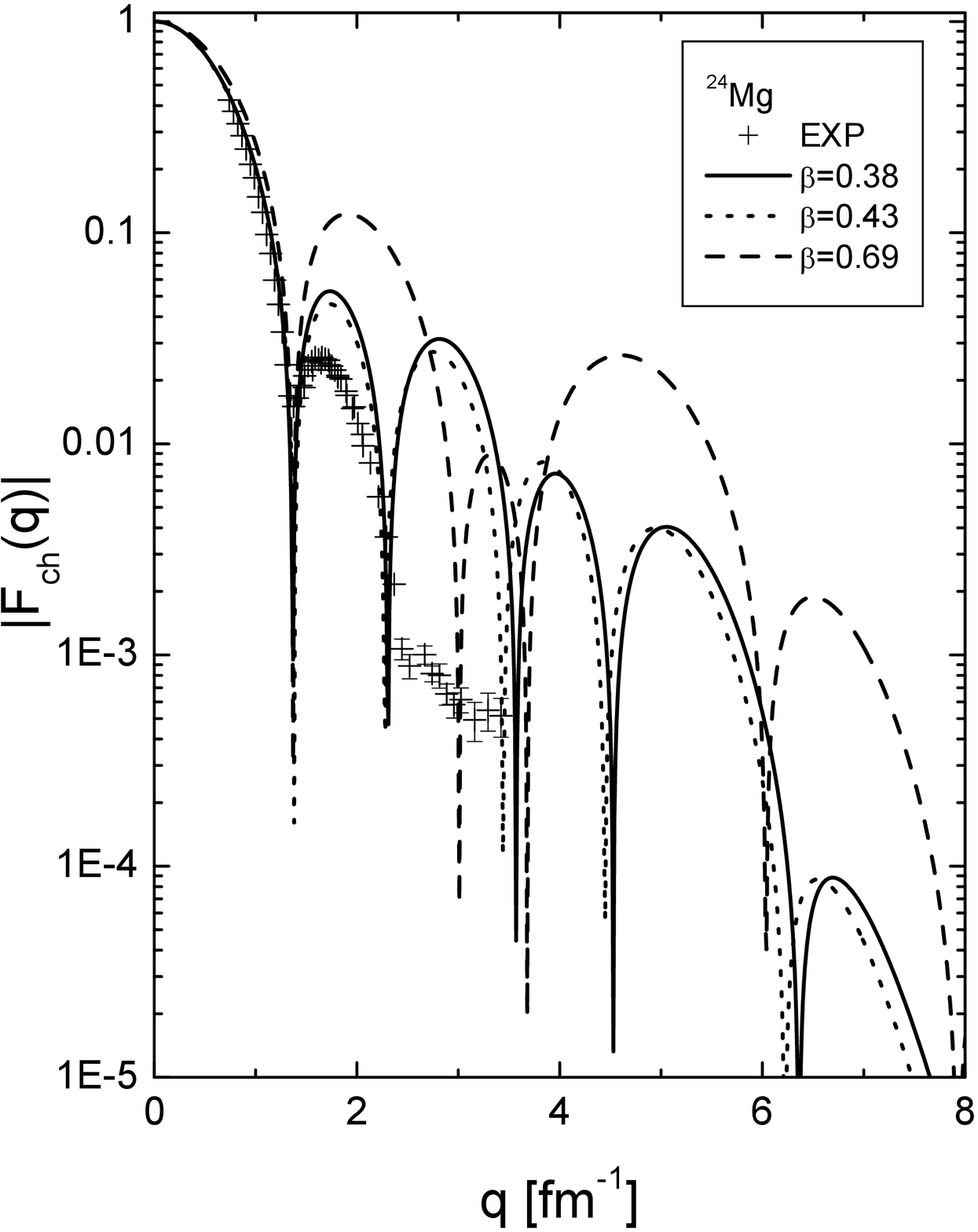}
\hspace{0.5cm}
 \includegraphics[height=8.5cm,width=8.5cm]{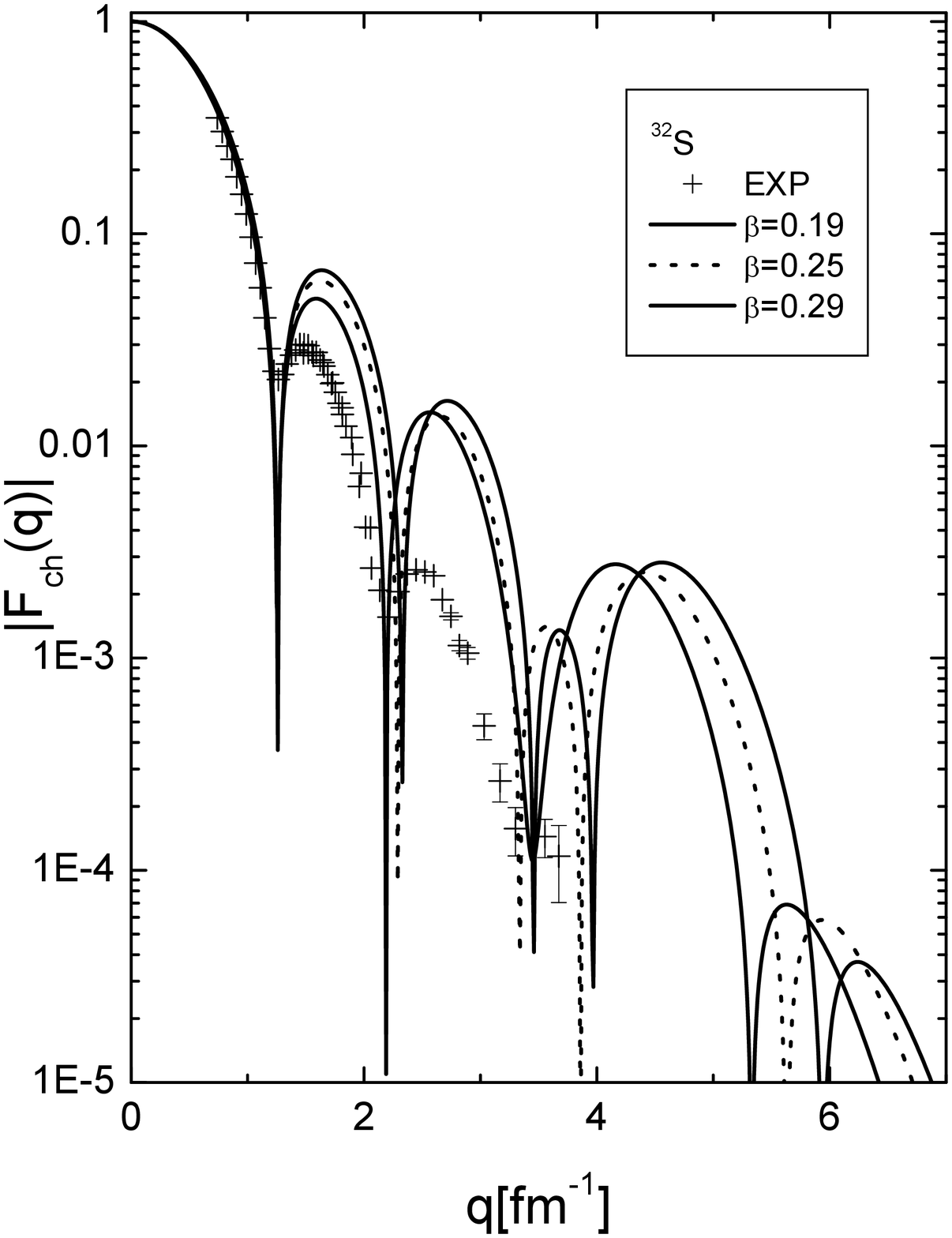}
 %\hspace{0.5cm}
  \caption{ Charge form factor of  $^{24}$Mg (up) and $^{32}$S
(down) for various values of the deformation parameter $\beta$
entering the deformed WS basis. Pairing correlations are also
taken into consideration through the BCS ground state. The values
of $\beta$ comes from Ref. \protect\cite{Stone01}. }
\end{figure}
%%%%%%%%%%%%%%%%%%%%%%%%%%%%%%%%%%%%

In the next part of our work, for each isotope studied we
investigated the influence of pairing correlations and deformation
on the details of the nuclear surface indicated by the occupancies
of the active orbits and the Fermi sea depletion (the sum of the
valence vacancy numbers, $U_{pi}^2$). In Fig. 3, the proton-orbit
occupancies versus the single particle energies are illustrated
(cases denoted as WS and WS+BCS) and in Table 2 (last column) we
report the Fermi sea depletions estimated for the case WS+BCS of
Fig. 3. From these results we conclude that, the largest depletion
corresponds to $^{24}$Mg nucleus (the most deformed isotope of our
set) and the smallest one to $^{32}$S (the least deformed isotope
of our set) which might be accidental. Concerning the two Mg
isotopes, evidently, the depletion of the less deformed $^{26}$Mg
is appreciably smaller than that of $^{24}$Mg showing the quite
strong effect of deformation on the smearing of the nuclear
surface. This result agrees with the "hardening" of the
valence-proton distribution found in Ref. \cite{Sound88} which is
due to the addition of two extra neutrons.

%%%%%%%%%%%%%%%%%%%%%%%%%%%%
%Fig3
\begin{figure}[h]
 \centering
 \includegraphics[height=13.5cm,width=13.5cm]{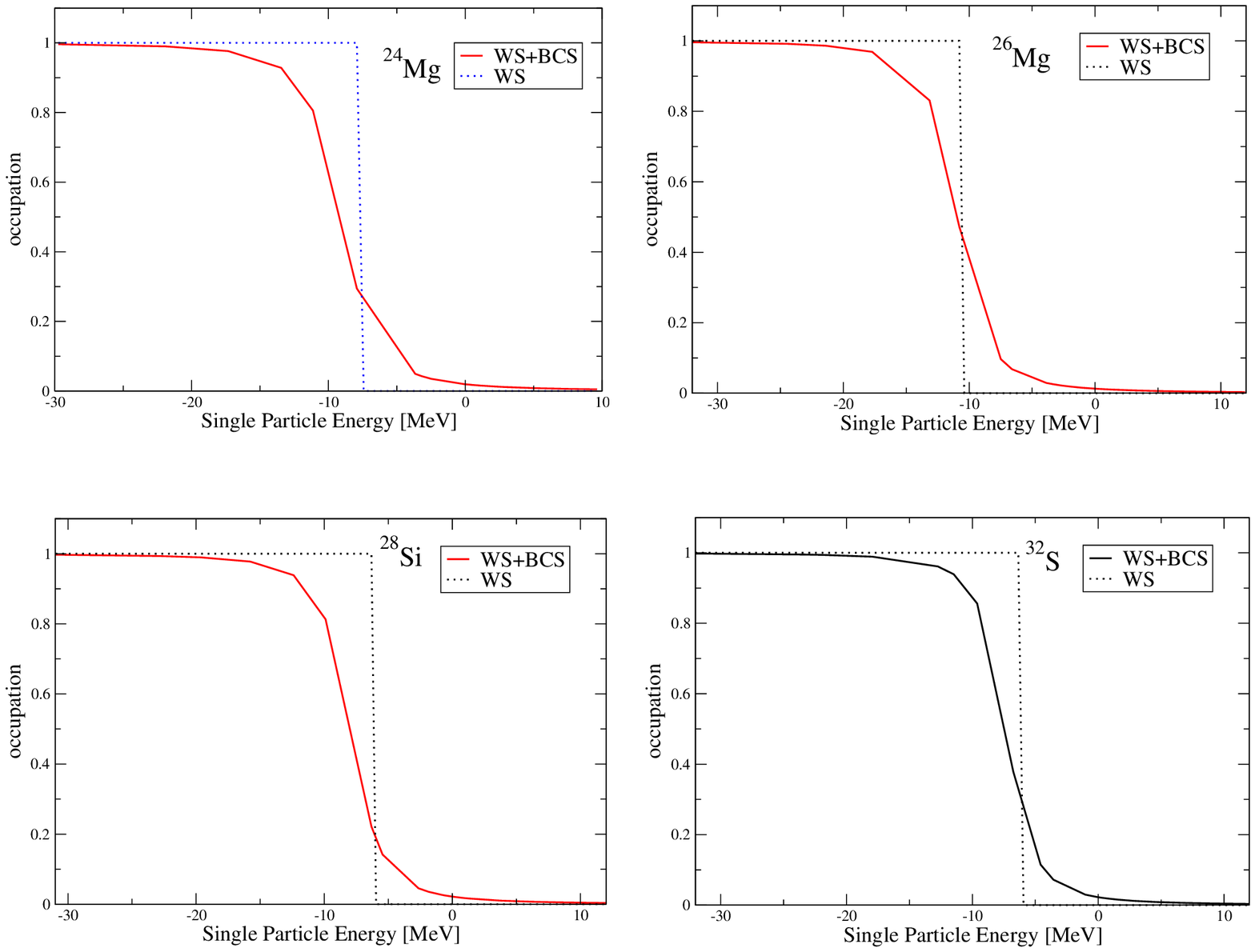}
 %\hspace{0.5cm}
  \caption{Proton occupation probabilities near the Fermi surface
  for the nuclei $^{24}_{12}$Mg,
$^{26}_{12}$Mg, $^{28}_{14}$Si and $^{32}_{16}$S. The diffuseness
of the Fermi surface reflects the pairing correlations and
deformation effects. The notation is the same with that of Fig. 1.
}
\end{figure}
%%%%%%%%%%%%%%%%%%%%%%%%%%%

In Table 2, for comparison, we quote the shell model Fermi sea
depletions of Refs. \cite{Sound88,Wessel97} discussed in
conjunction with elastic scattering data that provide information
about differences in occupations of individual orbits between the
pairs of neighboring even-even nuclei ($^{24}$Mg, $^{26}$Mg) and
($^{26}$Mg, $^{28}$Si). In this Table we also include the shell
model Fermi sea depletion calculations of Ref. \cite{Potbhare} and
the values extracted from the experimental data of Refs.
\cite{Pandya,Ishkhanov}. The authors of Ref. \cite{Potbhare},
using two well-tested effective two-body interactions
\cite{U-int,PW-int} and treating the above nuclei as spherically
symmetric systems, found drastically smaller depletions compared
to ours and appreciably smaller than the experimental ones and
those of Ref. \cite{Sound88}. From Table 2 one can infer that, the
experimental Fermi sea depletions of Ref. \cite{Pandya,Ishkhanov}
are overestimated by about 40-130 \% (depending on the nuclear
system) from our method, and they are underestimated by about
10-50 \% from the latter shell model calculations. The
calculations of Ref. \cite{Sound88}, in general, agree quite well
with the experimental results.
%%%%%%%%%%%%%%%%%%%%%%%%%%%%%%%%%%%%%%%%%%%%%%%%%%
\vspace{0.5 cm}
\begin{table}[h]
\begin{center}
\caption{ Fermi sea depletions for the deformed nuclear isotopes
$^{24}$Mg, $^{26}$Mg, $^{28}$Si and $^{32}$S evaluated by the
present work (last column) and previous spherical shell model
calculations. The experimental results are from the following
Refs. SM I \protect\cite{U-int}, SM II \protect\cite{PW-int}, SM
III \protect\cite{Wessel97}, Exp. I \protect\cite{Pandya} and Exp.
II \protect\cite{Ishkhanov}.
% The experimental results of Refs.
%\protect\cite{Pandya,Ishkhanov} are also quoted.
$^{\star}$ This
result comes from spherically symmetric Hartree-Fock calculations
discussed in Ref. \protect\cite{Wessel97} which include the
$0f_{7/2}$ orbital in the model space. }
\begin{tabular}{|cccccccc|}
 %\toprule
\hline \hline
&   &   &   &   &   &  &  \\
 Nucleus & $\Delta_{pp}$ (MeV) & SM I  & SM II
    & SM III
    & Exp. I & Exp. II  & This Work \\
\hline
&   &   &   &   &   &   &  \\
 $^{24}$Mg & 2.62 &  3.6  & 4.0 & 9.2 &  6.7  &  7.3  &  16.8  \\
&   &   &   &   &   &   & \\
 $^{26}$Mg & 2.52 &  4.3  & 3.9 & 6.7 &  5.8  &  6.0  &  12.6  \\
&   &   &   &   &   &   &  \\
 $^{28}$Si & 2.43 &  6.9  & 6.8 & 9.8 & 10.7  & 10.7  &  15.3  \\
&   &   &   &   &   &   & \\
 $^{32}$S  & 2.27 &  5.2  & 4.8 & 5.8$^{\star}$ &  5.0  &  5.6  &  11.1  \\
\hline \hline
%\botrule
\end{tabular}
\end{center}
\end{table}

The disagreement between these experimental data and our theory
might be related to the factors discussed below, after making the
following remarks: In our BCS treatment all orbits involved in our
rich model space are considered as active, whereas the shell model
results of Ref. \cite{Sound88,Potbhare} originate from the very
restricted model space $0d_{5/2}$$1s_{1/2}$$0d_{3/2}$ (both
interactions used in \cite{Potbhare} give rather similar orbit
occupancies even though they differ in their predictions of
occupancy dependent single particle energies, which means that for
each isotope, the corresponding to our Fig. 3 pictures given by
the latter two shell model calculations might be substantially
different compared to each other). The parameterization of the
deformed WS potential assumed in this paper, had originally been
proposed for spherical nuclei (ranging from $^{16}_{8}$O to
$^{208}_{82}$Pb). The confidence level of this parameterization
(and subsequently of our results for deformed nuclei) relies upon
its isospin-dependence characteristic which has indirectly been
checked from the reliability of descriptions done (under the same
assumptions with the present work) in Ref.
\cite{Sim-Mous,Sarrigu03,Sim-Pace}.

In order to elucidate further the relationship between the
deformation parameter $\beta$ and the depletion of the valence
shells, we have carried out a set of calculations with different
values of $\beta$ (originated from Ref. \cite{Stone01}, see also
Table 1) keeping fixed the pairing gap for the proton-proton and
neutron-neutron channel, $\Delta_{pp} = \Delta_{nn}$, in the BCS
procedure (restricting in this way as much as possible the
variation of pairing correlations effects). We have chosen the
nuclei $^{24}$Mg and $^{32}$S and the results are displayed in
Fig. 4. It is evident from this figure, that the Fermi sea
depletion decreseased  slightly as the deformation effects tend to
be canceled out and vice versa. More precisely the depletion
ranging in the case of $^{24}$Mg from $17 \%$ ($\beta=0.38$) to
$17.5 \%$ ($\beta=0.69$) and in case of $^{32}$S ranging from
$11.11 \%$ ($\beta=0.19$) to $11.31 \%$ ($\beta=0.29$). This is an
interesting result and may be accounted for as follows. By keeping
the pairing interaction rather constant, a decrease of the
deformation increases the orbit vacancies $U_{pi}^2$ and
subsequently causes more effective participation of the nuclear
core to the surface texture. Moreover, since for the latter
results (and those of Table 2) we have not considered other sorts
of correlations except pairing ones, the smearing of the Fermi
surface is artificially interpreted by increasing the vacancies of
the hole states (nuclear core), namely the Fermi surface
depletion. In the region of deformation parameter, $0.2
\le\beta\le 0.7$, where our method is applicable, the Fermi sea
depletion seems, for this reason, to be overestimated.  At this
point we should mention that, in the recent literature
\cite{Pandharipande97} authors assert that the depopulation of
states below the Fermi energy is much more pronounced than it was
believed in the past, since the quantitative information we have
today from both theory and experiment support occupation
probabilities in finite nuclei of the order of 75 \% only. Our
present results underly this result and it is expected from future
measurements on specific nuclei with higher precision and covering
larger range of momentum transfer to shed more light on the
situation.

%%%%%%%%%%%%%%%%%%%%%%%%%%%%
%Fig4
\begin{figure}[h]
 \centering
 \includegraphics[height=8.5cm,width=8.5cm]{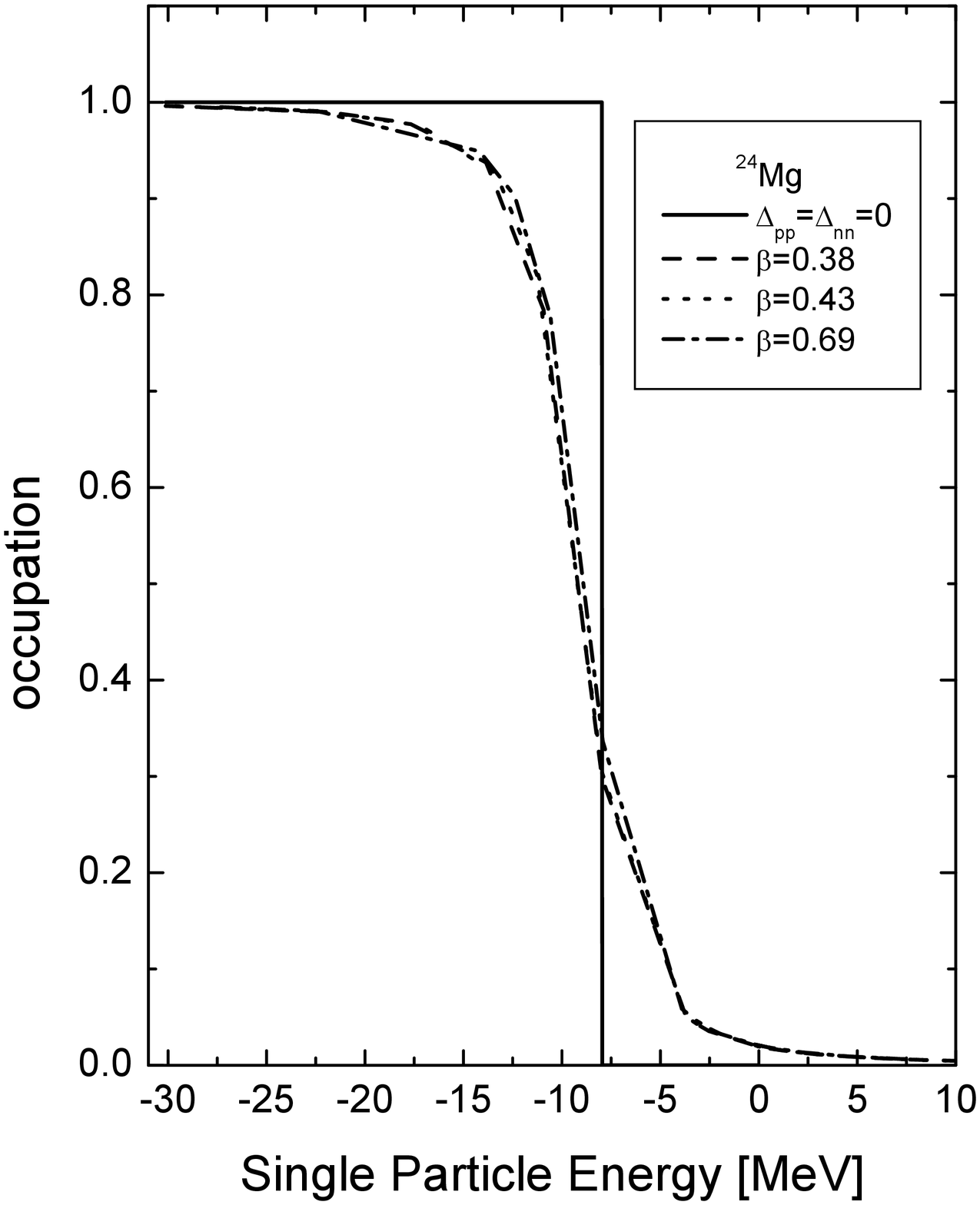}
\hspace{0.5cm}
\includegraphics[height=8.5cm,width=8.5cm]{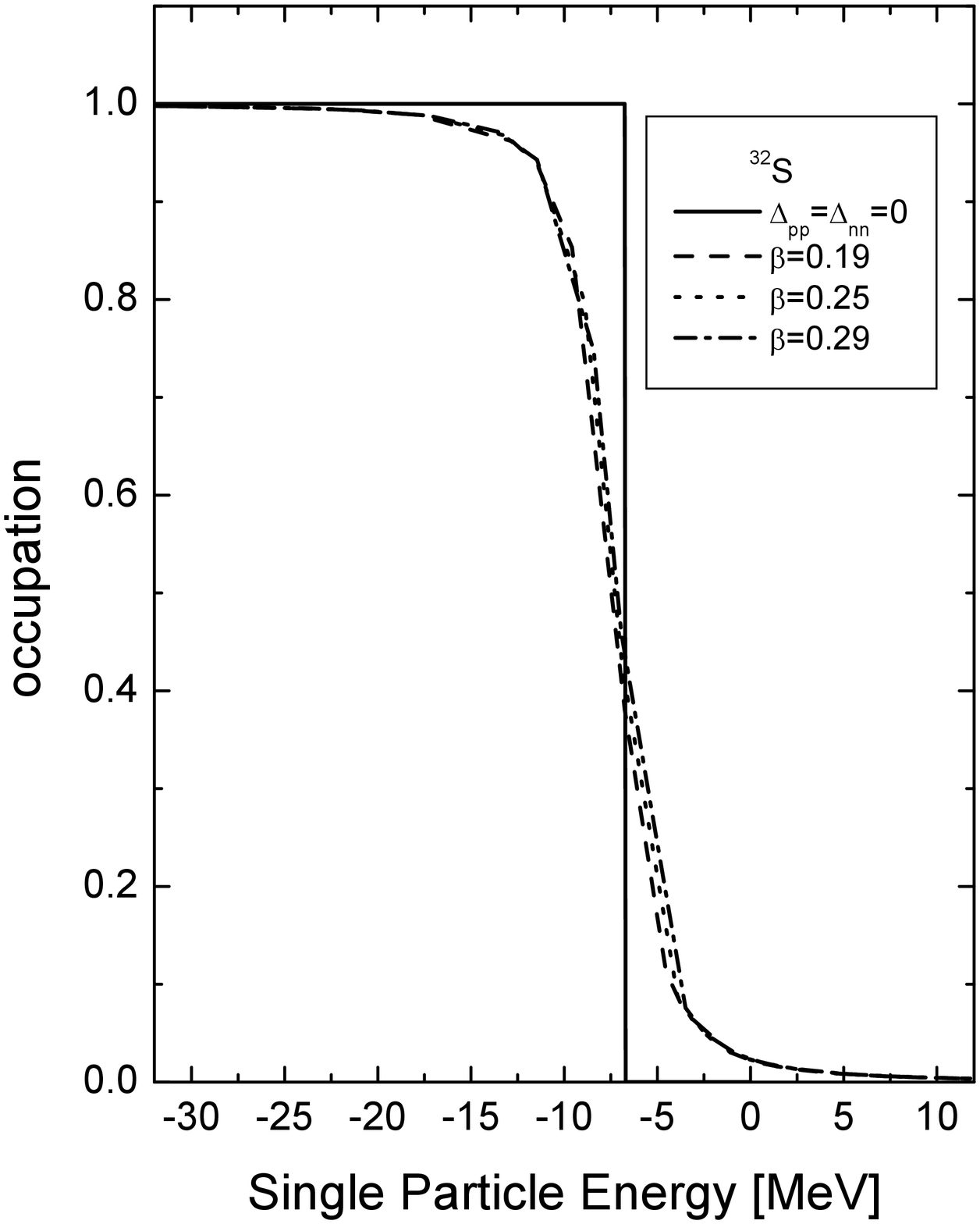}
 %\hspace{0.5cm}
 \caption{ The diffused Fermi surface and the Fermi sea depletion
 for $^{24}_{12}$Mg (up) and $^{32}$S (down)
produced by varying the deformation parameter $\beta$ in a
deformed Woods-Saxon basis (see Table 1). The strength for
proton-pair (neutron-pair) interaction in the BCS procedure is
fixed at the value $\Delta_{pp}=\Delta_{nn}=12.84/A^{1/2}$ MeV. }
\end{figure}
%%%%%%%%%%%%%%%%%%%%%%%%%%%

Before closing we note that, in our calculations, effects like
finite proton size, center-of-mass corrections, etc., have not
been taken into account. Instead, we restricted ourselves to the
evaluation of the point nucleons (protons) distribution, which is
equivalent to ignoring their internal substructure. This
approximation has extensively been used in nuclear structure
methods based on BCS ground state, because, the shape of the
charge form factor is mainly determined by the shape of the
point-proton density. In the present work we have also stayed
within the assumptions of the above approximation encouraged for
it by the rather well description of decay processes like the
double beta decay \cite{Sim-Pace} performed recently by using wave
functions constructed as described in Sect. 2. As it is well
known, the nuclear structure details affect considerably the
corresponding double beta decay rates since its transition matrix
elements are fairly sensitive.

\section{Summary and Conclusions }

In the present paper, at first, we formulated on the basis of BCS
quasi-particles a simultaneous consideration of deformation and
pairing correlations phenomena. The mean field description is
provided in terms of an axially symmetric Woods-Saxon potential
whereas for the two-body interaction we assumed schematic forces.
Next, we tested this method by studying gross ground state
properties like the nuclear form factor, the occupancy (vacancy)
numbers of the active orbitals, and the corresponding depletion of
the Fermi sea for the set of deformed nuclei, $^{24}_{12}$Mg,
$^{26}_{12}$Mg, $^{28}_{14}$Si and $^{32}_{16}$S. As it is well
known from nuclear structure calculations, the focus on such
primary gross properties, which is by itself a continuously
important subject of study, provides very good tests for
microscopic descriptions of the nuclear dynamics within many-body
theories. This is encouraged by the great resource of data for
these properties provided mainly by electron scattering
experiments.

From the conclusions extracted out of the present calculations, we
see that the use of deformed Woods-Saxon field in our BCS approach
reproduces quite well the experimental form factor data in the set
of isotopes studied, considerably better than the description
offered when utilizing a deformed harmonic oscillator basis. The
simultaneous consideration of the deformation and pairing
correlations affect largely the Fermi sea depletion and produces
appreciable smearing of the nuclear surface.

The investigation of the influence of the quadrupole deformation
(parameter $\beta$) on the surface textures of the above
moderately deformed nuclear systems, showed that the active orbit
occupancies and, subsequently, the depletion of the Fermi sea for
each isotope, are dependent slightly upon the parameter $\beta$.
The results found for the Fermi sea depletion are more than 40\%
(depending on the specific nuclear system) larger than those
emerging out of the old experimental data, from which one would
infer that our method overestimates, somehow, the Fermi sea
depletion through stronger deformation effects in order to
compromise for the correlation effects not considered by it.
However, recent information (experimental and theoretical) tend to
support larger depletions (of the order of 20-25 \%) which are
consistent with our present estimations. The remaining discrepancy
may be clarified from specific future experiments of greater
precision and larger momentum transfers.

\vspace{0.4 cm}

The authors would like to thank Professor I. Sick for providing us
with the experimental data for the charge form factor. One of the
authors (Ch.C.M) would like to thank Professor S.E. Massen for
useful comments and discussions. This work has been supported in
part (TSK) by the IKYDA-02 project.

% {\large \bf Acknowledgments}

\vspace{0.3 cm}

%%%%%%%%%%%%%%%%%%%%%%%%%%%%%%%%%%%%%%%%%%%%%%%%%%%%%%%%%%%%%%%%%%%%%%%%%%%%%

%%%%%%%%%%%%%%%%%%%%%%%%%%%%%%%%%%%%%%%%%%%%%%%%%%%%%%%%%%%%%%%%%%%%%%%%%%%%%
%%%%%%%%%%%%%%%%%%%%%%%%%%%%%%%%%%%%%%%%%%%%%%%%%%%%%%%%%%%%%%%%%%%%%%%%%%%%%%
%%%%%%%%%%%%%%%%%%%%%%%%%%%%%%%%%%%%%%%%%%%%%%%%%%%%%%%%%%%%%%%%%%%%%%%%%%%%%%

\end{document}